\documentclass[aps,nobibnote,pra,superscriptaddress,twocolumn,showpacs]{revtex4}
\usepackage{hyperref}
\usepackage[normalem]{ulem}
\usepackage{amsmath,amssymb}
\usepackage{graphicx}
\usepackage{feynmf}
\usepackage{array}
\usepackage{color}	
\usepackage{verbatim}	
\usepackage{MnSymbol}	
\newcommand{\rs}{\rm \scriptscriptstyle}

\newcommand{\A}{\mathcal{A}}

\newcommand{\T}{\mathcal{T}}

\begin{document}

\title{A tunable, nonlinear Hong-Ou-Mandel interferometer}

\author{D. Oehri}
\affiliation{Theoretische Physik,
  ETH Zurich, CH-8093 Zurich, Switzerland}

\author{M. Pletyukhov}
\affiliation{Institute for Theory of Statistical Physics 
  and JARA Ð-- Fundamentals of Future Information Technology, 
  RWTH Aachen, 52056 Aachen, Germany}

\author{V. Gritsev}
\affiliation{ Institute for Theoretical Physics, 
  University of Amsterdam, Science Park 904, 
  Postbus 94485, 1098 XH Amsterdam, The Netherlands}

\author{G. Blatter}
\affiliation{Theoretische Physik,
  ETH Zurich, CH-8093 Zurich, Switzerland}

\author{S. Schmidt}
\affiliation{Theoretische Physik,
  ETH Zurich, CH-8093 Zurich, Switzerland}

\date{\today}

\begin{abstract}
We investigate the two-photon scattering properties of a Jaynes-Cummings (JC)
nonlinearity consisting of a two-level system (qubit) interacting with a
single mode cavity, which is coupled to two waveguides, each containing a
single incident photon wave packet initially. In this scattering setup, we
study the interplay between the Hong-Ou-Mandel effect arising due to quantum
interference and effective photon-photon interactions induced by the presence
of the qubit.  We calculate the two-photon scattering matrix of this system
analytically and identify signatures of interference and interaction in the
second order auto- and cross-correlation functions of the scattered photons.
In the dispersive regime, when qubit and cavity are far detuned from each
other, we find that the JC nonlinearity can be used as an almost linear,
in-situ tunable beam splitter giving rise to ideal Hong-Ou-Mandel
interference, generating a highly path-entangled two-photon NOON state of the
scattered photons. The latter manifests itself in strongly suppressed waveguide
cross-correlations and Poissonian photon number statistics in each waveguide.
If the two-level system and the cavity are on resonance, the JC nonlinearity
strongly modifies the ideal HOM conditions leading to a smaller degree of path
entanglement and sub-poissonian photon number statistics.  In the latter
regime, we find that photon blockade is associated with bunched
auto-correlations in both waveguides, while a two-polariton resonance can lead
to bunched as well as anti-bunched correlations.
\end{abstract}


\maketitle

\section{Introduction}\label{sec:intro}
%
The Hong-Ou-Mandel (HOM) effect \cite{hong1987} lies at the heart of linear
optical quantum computing \cite{knill2001,kimble2008} and is utilized to test
the degree of indistinguishability of photons as well as the quality of single
photon sources.  Conventionally, the HOM effect is demonstrated as a
two-photon interference effect at a linear 50/50 beam splitter: when two
indistinguishable photon wave packets impinge on the beam splitter from two
different waveguides, they form an entangled two-photon NOON state with both
photons in the same output waveguide which produces a characteristic dip in
the coincidence probability of finding photons in both waveguides
simultaneously. This destructive interference is complete only for
indistinguishable photons with equal energy and zero time delay. Any deviation
from the ideal conditions leads to a diminishing of the dip, which thus
constitutes a measure for indistinguishability.

The HOM effect has been demonstrated experimentally using parametric down
conversion \cite{hong1987} and pulsed single photon sources operating at
optical frequencies \cite{santori2002, deriedmatten2003, legero2004,
beugnon2006, maunz2007}.  Recently, the HOM effect has attracted new attention
in the context of waveguide and circuit QED \cite{blais2004, wallraff2004,
longo2011, pletyukhov2014}, where it was demonstrated with unprecedented
precision for microwave photons as well \cite{lang2013, woolley2013}, using
recently developed microwave single photon sources and beam splitters. This
development paves the way for all-integrated linear optical quantum computing
at microwave frequencies.

In all experiments so far, the central element of the HOM effect, i.e., a
$50/50$ beam splitter, was based on a linear and static device, whose
single-photon transmission and reflection probabilities (which are constant
over a broad range of frequencies) are fixed by the manufacturing process.
From an experimental point of view it would be desirable to also utilize beam
splitters with in-situ tunable reflection and transmission probabilities.
\begin{figure}
\begin{center}
\includegraphics{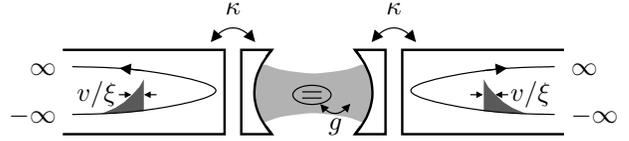}
\caption{
Schematic of the scattering setup with a qubit-cavity system interacting via
the light-matter interaction strength $g$ and directly connected to two
transmission lines giving rise to a cavity decay rate $\kappa=2\pi
g_\mathrm{w}^2$. We consider two single-photon wave packets with spatial width
$v/\xi$ impinging on the cavity from two different directions, e.g., emitted
by two independent single photon sources \cite{houck:07,lang:11}.
} 
\label{fig:setup} 
\end{center}
\end{figure}

In this paper we study theoretically two-photon scattering at a
Jaynes-Cummings nonlinearity \cite{shi:11}, i.e., a coupled qubit-cavity
system, connected directly to two transmission lines. Such a scattering setup
is readily realizable using state of the art circuit QED technology.  Hereby,
the transition frequency of the two-level system (qubit) is a flux-tunable
parameter, which allows to energetically shift the effective single-photon
resonances of the qubit-cavity system. Due to the finite width of the
resonances induced by the coupling to the transmission lines, one can
fine-tune the single photon transmission and reflection probabilities of the
scattering target. The Jaynes-Cummings nonlinearity may thus act as an in-situ
tunable beam splitter in a waveguide QED setup. This tunable beam splitter
works best for energies of the incoming photons with a bandwidth $\xi$ smaller
than the cavity decay rate $\kappa$. In circuit QED, this condition can be
satisfied by using a driven qubit-cavity system as a single photon source
\cite{houck:07,lang:11} which couples more weakly to the waveguide (leading
to sharp wave packets) than the beam splitter cavity.

In addition, the two-level system introduces a nonlinearity into the system,
which may modify the ideal HOM conditions.  Here, we investigate in detail the
interplay between the two-photon interference and effective photon-photon
interactions in a Jaynes-Cummings system. For this purpose, we utilize an
analytic scattering matrix approach recently developed for a two-level system
embedded in a chiral photonic waveguide \cite{pletyukhov2012}.  Based on this
analytic approach, we derive the exact two-photon scattering matrix of the
proposed setup and identify experimentally measurable signatures of the HOM
effect and effective photon-photon interactions in the second-order cross- and
auto-correlation functions of the two output modes. In the dispersive regime,
i.e., when the two-level system and the cavity are far detuned from each
other, the Jaynes-Cummings nonlinearity gives rise to a qubit-like and a
cavity-like scattering resonance.  In the vicinity of these resonances, the
two-photon scattering matrix of a qubit and a Kerr-like nonlinearity,
previously studied in the context of the HOM effect \cite{longo2011}, are
derived as limiting cases.

\section{Model}\label{sec:scaJC}

\begin{figure}
\begin{center}
\includegraphics{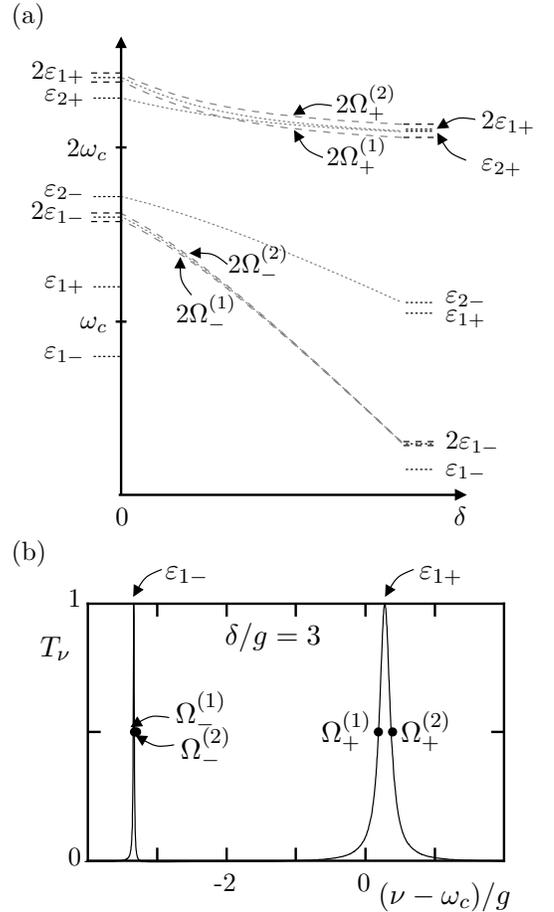}
\caption{
(a) Energy level diagram of the Jaynes-Cummings Hamiltonian showing the
upper/lower one- and two-polariton resonances $\varepsilon_{1\sigma}$ and
$\varepsilon_{2\sigma}$ defined in Eq.~(\ref{jcenergies}) for zero detuning
($\delta = 0$) on the left and for large positive detuning ($\delta > 0$) on
the right. Dashed lines correspond to the two-photon energies
$2\Omega^{(1,2)}_\sigma$ defined in Eq.~(\ref{diplocation}) leading to the
ideal HOM effect for sharp wave packets with $\xi\rightarrow 0$. Dotted lines
correspond to the two-polariton energies $\varepsilon_{2\sigma}$ and
$2\varepsilon_{1\sigma}$ as a function of detuning $\delta$. Their difference
measures the nonlinearity in the Jaynes-Cummings spectrum
$U=\varepsilon_{2\sigma}-2\varepsilon_{1\sigma}$. (b) The single-photon
transmission probability (i.e., transmission to the other waveguide)
$T_{\nu}=|t_\nu|^2$ exhibits two Lorentzian resonances located approximately
around the two-level system frequency $\omega_q$ and the cavity frequency
$\omega_c$ in the dispersive regime with $|\delta|=|\omega_c-\omega_q| \gg g$.
The energies $\Omega_{\sigma}^{(1,2)}$ correspond to ideal HOM conditions with
$T_{\nu}=R_{\nu}=1/2$ (where $R_{\nu}=|r_\nu|^2$ denotes the reflection
probability and we have chosen $\kappa/g=0.1$).
}
\label{fig:fig2new}
\end{center}
\end{figure}

Figure \ref{fig:setup} shows the scattering setup considered in this paper.
The total Hamiltonian of this system is given by
\begin{align} 
	H = H_{\rm\scriptscriptstyle JC}+H_{\rm w}+V 
\end{align} 
with the Jaynes-Cummings Hamiltonian 
\begin{align}
	H_{\rm\scriptscriptstyle JC} = \omega_c b^\dagger b+ \omega_q\sigma^{+}\sigma^{-} 
							+g (b^\dagger \sigma^{-}+b \sigma^{+}) 
\end{align} 
describing the coupled qubit-cavity system, we set $\hbar=1$.  The Hamiltonian
$H_{\rm w}$ of the transmission lines describes photons in chiral states (see
Fig.~\ref{fig:setup}) with linear dispersion $\nu=vk$ with wavevector $k>0$
and velocity $v$, i.e.,
\begin{align} 
	H_{\rm w}=\sum_{i=1,2}\int d\nu\, \nu\, a_{i\nu}^\dagger a_{i\nu}\,, 
\end{align} 
which linearly couple to the scatterer via
\begin{align} 
	V=g_\mathrm{w} \sum_{i=1,2}\int d\nu\,(b^\dagger a_{i\nu}+\text{h.c.})\,.
\end{align} 
Here, we have introduced the cavity photon operator $b$ and the operators
$a_{i\nu}$ describing photons with energy $\nu$ in waveguide $i$.  The
two-level system, described by the raising/lowering operators for a two-level
system $\sigma^\pm$, introduces a nonlinearity into the system due to the
light-matter coupling $\sim g$.  In our model, we neglect a coupling of the
two-level system to other modes of the system and environment, because the
associated spontaneous emission rates can be engineered to be several orders
of magnitude smaller than the qubit-cavity coupling $g$ and cavity decay rate
$\kappa=2\pi g_\mathrm{w}^2$ to the leads (for typical parameter values in
circuit QED systems, see Ref.~\cite{schmidt:13}).

The scattering resonances are determined by the eigenstates of the
Jaynes-Cummings Hamiltonian: The ground state
$|\psi_{0}\rangle=|0,\downarrow\rangle$ consists of zero photons in the cavity
and the two-level system in its ground state with energy $\varepsilon_{0}=0$.
The excited states ($n>0$) are
\begin{align}
\label{jcstates}
|\psi_{n+}\rangle&=\cos\frac{\theta_n}{2} |n-1,\uparrow\rangle  
			   +\sin\frac{\theta_n}{2}    |n,\downarrow\rangle,\\
|\psi_{n-}\rangle&=-\sin\frac{\theta_n}{2} |n-1,\uparrow\rangle
			   +\cos\frac{\theta_n}{2} |n,\downarrow\rangle,
\end{align}
with the angle $\tan\theta_n=- 2g\sqrt{n}/\delta$ and the detuning $\delta =
\omega_c - \omega_q$. They correspond to $n$ excitations, i.e., polariton
quasi-particles, which form a superposition of the state
$|n,\downarrow\rangle$ with $n$ photons in the cavity and the two-level system
in the ground state and the state $|n-1,\uparrow\rangle$ with $(n-1)$ photons
in the cavity and the two-level system in the excited state. The corresponding
energies are given by
\begin{align}
\label{jcenergies}
\varepsilon_{n\pm}=n\omega_c-\delta/2  \pm \sqrt{(\delta/2)^2+ ng^2}.
\end{align}
The JC eigenenergies give rise to single-photon scattering resonances with
Lorentzian peaks at $\varepsilon_{1\pm}$ as a function of incoming photon
energy.  Due to the coupling to the waveguide, the cavity states also attain a
finite lifetime $\sim 1/\kappa$ with $\kappa=2\pi g_\mathrm{w}^2$.  In the
next section, we will derive explicit expressions for the single-photon and
two-photon scattering matrix of this system.

\section{Scattering formalism}
We consider the generic situation, where an initial one- or two-photon state
$|\psi_{\rm in}\rangle$ is prepared inside the wave-guides at $t_{\rm
in}=-\infty$, interacts (scatters) at the Jaynes-Cummings nonlinearity and is
observed as the out-going state $|\psi_{\rm out}\rangle$ in a photon detector
at $t_{\rm out}=+\infty$.  The relation between the initial and the scattered
state is given by the $S$-matrix, i.e.,
\begin{align}
	|\psi_{\rm out}\rangle=S|\psi_{\rm in}\rangle\quad{\rm with}\quad S
	=\T \exp\bigl[-i\int_{-\infty}^{\infty}V(t)dt\bigr]\,.
\end{align}
Here, we have written the coupling Hamiltonian in the interaction
representation, i.e., $V(t)=\exp(i H_0 t) V \exp(-i H_0t)$ with
$H_0=H_{\rm\scriptscriptstyle JC}+H_{\rm w}$.

In the following, we will need the one-photon $S$-matrix (for the scattering
at an empty cavity/qubit, i.e., void of excitations) written in a Fock state
basis as
\begin{align}
\label{spmatrix}
P_0 S^{(1)} P_0 = P_0 \sumint S^{i j}_{\nu^\prime \nu}\, a^\dagger_{i\nu^\prime} a_{j\nu} 
\end{align}
with matrix elements $S_{\nu^\prime\nu}^{ij}= \langle 0 |
a_{i\nu^\prime}S^{(1)}a^{\dagger}_{j\nu} | 0 \rangle$. Here $|0 \rangle \equiv
|\psi_0 \rangle \otimes |0\rangle_{\mathrm{w}}$ is the state void of
excitations, i.e., no photons in cavity and waveguide, and the two-level
system residing in its ground state; $P_0 = |\psi_0 \rangle \langle \psi_0|$
is the projector onto the dark state of the cavity, i.e., $|\psi_0\rangle=|0,
\downarrow\rangle$. The symbol $\sumint = \sum_{i j}\int_{\nu^\prime \nu}$
denotes summation (integration) over all arabic (greek) indices.  Similarly,
the two-photon scattering matrix is defined as
\begin{align}
\label{tpmatrix}
P_0 S^{(2)} P_0 =
\frac{P_0}{4}\sumint_1 \sumint_2 S^{i_1 i_2 j_1 j_2}_{\nu^\prime_1\nu^\prime_2\nu_1\nu_2}\, 
a^\dagger_{i_1\nu^\prime_1} a^\dagger_{i_2\nu^\prime_2} a_{j_1\nu_1} a_{j_2\nu_2},
\end{align}
with $S^{i_1i_2j_1j_2}_{\nu^\prime_1\nu^\prime_2\nu_1 \nu_2}=
\langle 0 | a_{i_2\nu^\prime_{2}}a_{i_1\nu^\prime_{1}}
S^{(2)}a^{\dagger}_{j_1\nu_1}a^{\dagger}_{j_2\nu_2} | 0 \rangle$.

Both, single- and two-photon scattering matrix elements are calculated in
App.~\ref{app:sca-ma} by making use of the formalism developed in
Ref.~\onlinecite{pletyukhov2012} \footnote{We extend the integration range in
all energy integrals to $\nu\in(-\infty,\infty)$ which is a good approximation
as long as the frequency of the incoming photons $\nu_0>0$ (comparable to the
resonator and two-level system frequencies) sets the largest energy scale in
the problem.}.  For the single photon matrix elements in (\ref{spmatrix}) we
obtain
\begin{align}
\label{selements1}
S^{11}_{\nu^\prime\nu}&=S^{22}_{\nu^\prime\nu}=r_{\nu} \delta(\nu^\prime-\nu),\\
\label{selements2}
S^{12}_{\nu^\prime\nu}&=S^{21}_{\nu^\prime\nu}=t_{\nu}\delta(\nu^\prime-\nu),
\end{align}
with the reflection (= scattering to the same waveguide) amplitude 
\begin{align}
r_\nu=\frac{(\nu-\omega_q)(\nu-\omega_c)-g^2}{(\nu-\omega_q)(\nu-\omega_c+i\kappa)-g^2}
\label{eq:r}
\end{align}
and the transmission (= scattering to the other wave\-guide) amplitude 
\begin{align}
t_\nu=\frac{-i\kappa(\nu-\omega_q)}{(\nu-\omega_q)(\nu-\omega_c+i\kappa)-g^2}\,.
\label{eq:t}
\end{align}
The reflection amplitude vanishes for $\nu=\varepsilon_{1,\pm}$, giving rise
to resonances in transmission, see Fig.~\ref{fig:fig2new}(b). The one-photon
resonance energies and widths are determined by $\tilde{\varepsilon}_{1,\pm}$
which is obtained from $\varepsilon_{1,\pm}$, cf. Eq.~\eqref{jcenergies}, by
replacing the cavity frequency $\omega_c$ by
$\tilde{\omega}_c=\omega_c-i\kappa$  (and correspondingly replacing the
detuning $\delta$ by $\tilde{\delta}=\delta-i\kappa$; the $n$-excitation
resonances $\tilde{\varepsilon}_{n,\pm}$ are defined in the same way). For
large positive detuning $\delta\gg g$, the upper polariton resonance at
$\varepsilon_{1,+}$ is cavity-like with a width $\kappa$, while the lower
polariton resonance at $\varepsilon_{1,-}$ is qubit-like with an effective
width $\kappa g^2/\delta^2$.

In the context of the Hong-Ou-Mandel effect it is also useful to define the
single-photon energies
\begin{align}
\label{diplocation}
\Omega^{(1,2)}_\sigma&= \frac{1}{2}\Bigl(\omega_c + \omega_q \pm \kappa 
					+ \sigma \sqrt{(\delta\pm\kappa)^2+4g^2}\Bigr)
					\,, \quad \sigma=\pm\,,
\end{align}
where the single-photon transmission and reflection probabilities are equal to
$1/2$, corresponding to a 50/50 beam splitter operation, see
Fig.~\ref{fig:fig2new}(b).

The two-photon scattering matrix in (\ref{tpmatrix}) is of the form
\begin{align}
\label{telements}
S^{i_1i_2j_1j_2}_{\nu^\prime_1\nu^\prime_2\nu_1\nu_2}=
S^{i_1j_1}_{\nu^\prime_1\nu_1}S^{i_2j_2}_{\nu^\prime_2\nu_2}
+
S^{i_2j_1}_{\nu^\prime_2\nu_1}S^{i_1j_2}_{\nu^\prime_1\nu_2}
+
i\T^{(2)}_{\nu^\prime_1\nu^\prime_2\nu_1\nu_2}\,.
\end{align}
The first two terms describe two uncorrelated single photon scattering events
as obtained from Eqs.~\eqref{selements1} and~\eqref{selements2}. The third
term on the r.h.s of Eq.~(\ref{telements}) describes the non-trivial
two-photon scattering process described by the $T$-matrix element
\begin{align}
\label{tmelement}
\T^{(2)}_{\nu^\prime_1\nu^\prime_2\nu_1 \nu_2}&=
\delta(\nu^\prime_1+\nu^\prime_2-\nu_1-\nu_2)
\frac{\kappa^2g^4}{\pi}
\frac{(\nu_1+\nu_2-\tilde{\varepsilon}_{1,+}-\tilde{\varepsilon}_{1,-})}
{\prod_{\alpha=\pm}(\nu_1+\nu_2-\tilde{\varepsilon}_{2,\alpha})}
\nonumber\\
&\quad\times\frac{\prod_{\alpha=\pm}(\nu_1+\nu_2-2\tilde{\varepsilon}_{1,\alpha})}
{\prod_{i=1,2}\prod_{\alpha=\pm}(\nu^\prime_i-\tilde{\varepsilon}_{1,\alpha})(\nu_i-\tilde{\varepsilon}_{1,\alpha})}\,.
\end{align}
Note that the $T$-matrix elements do not depend on the waveguide indices due
to the symmetric coupling of the two transmission lines.  The results in
(\ref{spmatrix})-(\ref{tmelement}) are consistent with those derived in
Ref.~\cite{shi:11} (using a different approach) and enable us to calculate the
wavefunction as well as photon statistics in the output modes for arbitrary
incoming states containing up to two photons. In the next section we apply
these results to the specific situation depicted in Fig.~\ref{fig:setup} with
two single-photon wave packets approaching the cavity in different waveguides.

\section{Hong-Ou-Mandel effect}\label{sec:HOM}
The Hong-Ou-Mandel effect \cite{hong1987} is a two-particle-interference
effect describing the spatial bunching of two indistinguishable photons which
arrive at the same time at an ideal 50/50 beam splitter from two different
incoming arms and always end up in the same outgoing arm. The effect is
fundamentally related to the bosonic exchange properties of the photons.  In
the following, we will discuss such a Hong-Ou-Mandel (HOM) effect for two
photon scattering at a JC nonlinearity.  As discussed above, the single-photon
scattering characteristics of a JC nonlinearity provide equal transmission and
reflection probability at energies $\Omega^{(1,2)}_{\pm}$ (around the two
polariton energies $\varepsilon_{1,\pm}$) such that the cavity-qubit system
may act as an ideal 50/50 beam splitter. However, due to the finite width of
the photon wave packets, the ideal beam splitter conditions can only be
satisfied for one spectral component of the wave packets. Furthermore, the
nonlinearity of the JC system induced by the coupling of the photons to the
two-level system modifies the two-photon scattering properties compared to a
linear 50/50 beam splitter. Both finite wave packet width and nonlinearity may
wash out the quantum interference leading to the HOM effect and thus will be
investigated in detail in the following.

In order to investigate the HOM effect, we consider two photons incoming in
different waveguides described by the (normalized) incoming state
\begin{align}
|\psi_{\rm in}\rangle=
\iint d\nu_1  d\nu_2
f^{\rs (1)}_{\nu_1} f^{\rs (2)}_{\nu_2} 
a^\dagger_{1\nu_1}
a^\dagger_{2\nu_2}| 0 \rangle
\label{eq:in_state}
\end{align}
with the function $f^{\scriptscriptstyle (i)}_{\nu}$ describing the incoming photon wave
packet in lead $i=1,2$, satisfying $\int d\nu |f_{\nu}^{(i)}|^2=1$.  The
outgoing state after scattering is related to the incoming state via
$|\psi_{\rm out}\rangle=S|\psi_{\rm in}\rangle$ which leads to
\begin{align}
\label{eq:outstate}
|\psi_{\rm out}\rangle
&=|\psi^{11}_{\rm out}\rangle+|\psi^{22}_{\rm out}\rangle+|\psi^{12}_{\rm out}\rangle\\
&=
\iint d\nu_1 d\nu_2
f^{\rs (1)}_{\nu_1}
f^{\rs (2)}_{\nu_2} \nonumber\\
&\qquad\times\Biggl( \frac12 \iint d\nu_1^\prime
 d\nu_2^\prime
S^{11,12}_{\nu^\prime_1\nu^\prime_2,\nu_1\nu_2}
a^\dagger_{1\nu^\prime_1}
a^\dagger_{1\nu^\prime_2}| 0 \rangle\nonumber\\
&\hspace{30pt}+\frac12 \iint d\nu_1^\prime
 d\nu_2^\prime
S^{22,12}_{\nu^\prime_1\nu^\prime_2,\nu_1\nu_2}
a^\dagger_{2\nu^\prime_1}
a^\dagger_{2\nu^\prime_2}|0 \rangle\nonumber\\
&\hspace{30pt}+\iint d\nu_1^\prime
 d\nu_2^\prime
S^{12,12}_{\nu^\prime_1\nu^\prime_2,\nu_1\nu_2}
a^\dagger_{1\nu^\prime_1}
a^\dagger_{2\nu^\prime_2}| 0 \rangle\Biggr)
\nonumber
\end{align}
with three contributions; the first two contributions describe two photons
scattered into the same waveguide and the third contribution describes the
scattering into different waveguides. The HOM effect arises if the latter
contribution $|\psi_{\rm out}^{12}\rangle$ vanishes. We thus define a HOM
parameter
\begin{eqnarray}
\label{defgamma}
\gamma= \langle \psi_{\rm out} | n_1 n_2 | \psi_{\rm out} \rangle =  \langle\psi_{\rm out}^{12}|\psi_{\rm out}^{12}\rangle \,,
\end{eqnarray}
with $n_i=\int d\nu a_{i\nu}^\dagger a_{i\nu}$ which yields the coincidence
probability of finding one photon in the left and one photon in the right
waveguide. Thus, $\gamma=0$, corresponds to a superposition of two states, one
with both photons in the left and one with both photons in the right
waveguide, i.e., a two photon entangled NOON state with respect to the
waveguide degrees of freedom. Using Eq.~\eqref{eq:outstate}, we obtain from
(\ref{defgamma})
\begin{align}
\gamma
&=
\iint d\nu_1^\prime d\nu_2^\prime
\Bigl|
\iint d\nu_1 d\nu_2
S^{12,12}_{\nu^\prime_1\nu^\prime_2,\nu_1\nu_2}
f^{\rs (1)}_{\nu_1}
f^{\rs (2)}_{\nu_2}
\Bigr|^2.
\label{eq:gammaHOMgen}
\end{align}
In the following, we will consider Lorentzian wave packets described by
\begin{eqnarray}
f^{\scriptscriptstyle (i)}_{\nu}=\sqrt{\frac{\xi}{2\pi}} \frac{e^{-i\nu (t-t_i)}}{\nu-\nu_{0i}+i\xi/2} 
\label{eq:wavepacket}
\end{eqnarray}
around energies $\nu_{0i}>0$ with width $\xi$. The times $t_i$ correspond to
the instants when the fronts of the wave packets reach the mirrors of the
cavity, such that the time delay between them is given by $\Delta t=t_1-t_2$.
The integrals in Eq.~\eqref{eq:gammaHOMgen} can be calculated analytically
which results in explicit algebraic expressions for the HOM parameter, which,
however, are rather lengthy and thus have been omitted here for brevity.
\begin{figure}[t]
\centering
\includegraphics[width=0.49\textwidth,clip]{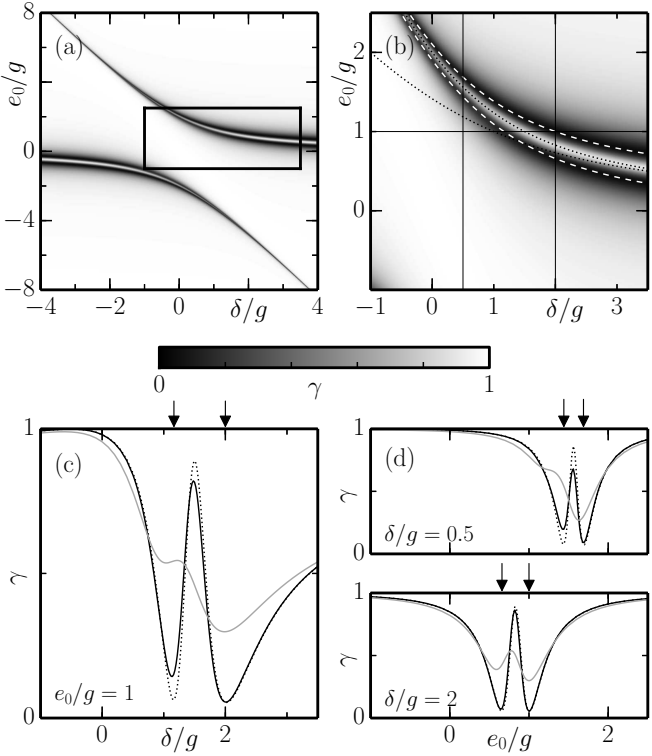}
\caption{\label{fig2} 
(a) HOM parameter $\gamma$ (defined in Eq.~(\ref{defgamma})) as a function of
the detuning $\delta$ and the two-photon energy
$e_0=\nu_{01}+\nu_{02}-2\omega_c$ (defined relative to the cavity frequency)
for $\kappa/g=0.1$ and $\xi/\kappa=0.1$.  (b) Inset in (a) (black box)
together with the two photon energies depicted in Fig.~\ref{fig:fig2new}(a)
(dashed, $2\Omega_{+}^{(1,2)}$, and dotted lines, $2\varepsilon_{1,+}$ and
$\varepsilon_{2,+}$, with the same notation as in Fig.~\ref{fig:fig2new}(a)).
Thin horizontal and vertical lines correspond to the cuts in (c) and (d), see
below. (c) Cuts through (b) at constant two-photon energy $e_0/g=1$ for wave
packets of width $\xi/\kappa=0.1$ (black) and $\xi/\kappa=1$ (light gray
lines). The dotted lines correspond to the linear approximation (for
$\xi/\kappa=0.1$), where correlation contributions due to the $T$-matrix
elements in Eq.~(\ref{tcontrib}) are neglected. Vertical arrows correspond to
the detuning for which $\nu_{01}+\nu_{02}=2\Omega^{(1,2)}_+$ (ideal HOM
conditions). (d) Cuts through (b) at constant detuning $\delta=0.5g$ (upper
panel) and $\delta=2g$ (lower panel).
}
\vspace{-0cm}
\end{figure}

Figure \ref{fig2} shows the behavior of the HOM parameter $\gamma$ as obtained
from the exact calculation of Eq.~\eqref{eq:gammaHOMgen} for $\Delta
t=\nobreak0$ and equal energies of the incoming photons $\nu_{01}=\nu_{02}$.
Before discussing these results in detail, it is instructive to separate
uncorrelated and correlated contributions in Eq.~\eqref{eq:gammaHOMgen}.  From
the result for the two-photon scattering matrix in (\ref{telements}), we can
rewrite
\begin{eqnarray}
\label{exactgamma}
\gamma=\iint d\nu_1^\prime d\nu_2^\prime 
|\mathcal{A}^{12}_{\nu^\prime_1,\nu^\prime_2}|^2
\end{eqnarray}
with
\begin{eqnarray}
\label{defA}
\mathcal{A}^{12}_{\nu^\prime_1,\nu^\prime_2}=
\mathcal{A}^{12,\rm lin}_{\nu^\prime_1,\nu^\prime_2} +\mathcal{C}_{\nu^\prime_1,\nu^\prime_2}
\end{eqnarray}
where 
\begin{eqnarray}
\mathcal{A}^{12,\rm lin}_{\nu^\prime_1,\nu^\prime_2}=
t_{\nu_1^\prime}t_{\nu_2^\prime}f^{\rs (1)}_{\nu_1^\prime}f^{\rs (2)}_{\nu_2^\prime}
+r_{\nu_1^\prime}r_{\nu_2^\prime}f^{\rs (1)}_{\nu_2^\prime}f^{\rs (2)}_{\nu_1^\prime}
\label{Alin}
\end{eqnarray}
and
\begin{eqnarray}
\label{tcontrib}
\mathcal{C}_{\nu^\prime_1,\nu^\prime_2}=i
\iint d\nu_1 d\nu_2 \T^{(2)}_{\nu^\prime_1\nu^\prime_2,\nu_1\nu_2}f^{\rs (1)}_{\nu_1}f^{\rs (2)}_{\nu_2}.
\end{eqnarray}
In the linear approximation, we neglect the contributions in
\eqref{exactgamma} arising from the $T^{(2)}$-matrix. Assuming furthermore
infinitely sharp wave packets $\xi \ll \kappa$ as well as zero initial energy
detuning, i.e., $\nu_{01}=\nu_{02}=\nu_{0}$, and zero time delay $\Delta t=0$,
we can write the HOM parameter in a simplified form as
\begin{align}
\label{eq:linresult}
\gamma_{\rm lin}&=\iint d\nu_1^\prime d\nu_2^\prime 
|\mathcal{A}^{12,\rm lin}_{\nu^\prime_1,\nu^\prime_2}|^2\nonumber\\
&=\prod_{\sigma=\pm} \frac{(2\nu_0-2\Omega^{(1)}_{\sigma})^2 (2\nu_0-2\Omega^{(2)}_{\sigma})^2} 
{    |2\nu_0-2\tilde{\varepsilon}_{1,\sigma}|^4 }
\end{align}
with $\Omega^{(1,2)}_\sigma$ defined in Eq.~\eqref{diplocation}. As expected,
the result of the linear approximation in (\ref{eq:linresult}) predicts
perfect HOM-like interference with $\gamma_{\rm lin}=0$ for infinitely sharp
wave-packets at the single-photon energies
$\nu_{01}=\nu_{02}=\Omega^{(1,2)}_\sigma$ where ideal 50/50 beam splitter
conditions prevail, see Fig.~\ref{fig:fig2new}.

Figure \ref{fig2}(a) shows $\gamma$ for non-ideal but favorable conditions
$\kappa/g=0.1$ and $\xi/\kappa=0.1$.  The energies $\Omega^{(1,2)}_{\sigma}$,
located around the one-polariton energy $\varepsilon_{1,\sigma}$, give rise to
a double dip feature in Fig.~\ref{fig2}(a) [cf. zoom in Fig.~\ref{fig2}(b) and
cuts in Fig.~\ref{fig2}(d)].  For large detuning $\delta\gg g$, we find two
well separated (approximately by $\kappa$) dips for positive (negative)
detuning around the upper (lower) polariton resonance. For negative (positive)
detuning the two HOM features around the upper (lower) polariton resonance are
very sharp and separated by $\kappa g^2/\delta^2$ only. The two dips merge
into one dip due to finite wave packet width for $\xi\lesssim\kappa
g^2/\delta^2$ and are completely washed out for $\xi>\kappa g^2/\delta^2$.
Figure~\ref{fig2}(b) shows a good agreement between the exact result as
obtained from \eqref{eq:gammaHOMgen} and the dip location predicted from the
linear approximation as discussed above.  In fact, the outgoing state with the
highest degree of path-entanglement (smallest value of $\gamma$) is always
found for the ideal HOM conditions (dashed lines in Fig.~\ref{fig2}(b))
independent of the detuning.  The overall degree of entanglement, however,
becomes maximal ($\gamma \approx 0$) only in the dispersive regime.
Additionally, one observes a dipole-induced-transparency like (DIT like)
effect \cite{waks2006} with $|t_\nu| \approx 1$ if the photons are tuned into
resonance with the one-polariton state $2\nu_0\approx 2\varepsilon_{1+}$
(upper dotted line in \ref{fig2}(b)) leading to an almost non-entangled
out-state with one photon in each waveguide similar to the incoming state
($\gamma\approx 1$).

Figures \ref{fig2}(c) and \ref{fig2}(d) represent cuts through the inset in
\ref{fig2}(b) at fixed energy (c) and detuning (d) of the incoming photons.
Both parameters can be used to tune in or out of the HOM-like interference.
Correlation effects attributed to the difference between solid and dotted
lines in \ref{fig2}(c) and \ref{fig2}(d) come into play for small detuning and
wash out the HOM effect, but are irrelevant in the dispersive regime with
moderately large detuning.  We also note that for broad wave packets with
$\xi\sim\kappa$ (gray curve) the HOM interference is washed out independent of
the detuning. This is due to the fact that ideal beam splitter conditions with
50/50 transmission/reflection probability can only be achieved for infinitely
sharp wave packets with $\xi \rightarrow 0$, otherwise not all parts of the
wave packet are scattered with equal probabilities as mentioned at the
beginning of this section.

The reason for the different behaviour at positive (negative) detuning in
Figure~\ref{fig2} is due to the change of the qubit/photon nature of the
one-polariton resonance: with increasing positive (negative) detuning, the
lower (upper) polariton resonance becomes more qubit (photon) - like and is
thus strongly decoupled from (coupled to) the photon scattering and
interference process, which leads to the HOM effect.  More specifically, in
the strongly dispersive limit corresponding to large positive detuning
$\delta\gg g$ with $\theta_n \rightarrow \pi^-$ in \eqref{jcstates}, the
polariton state $|\psi_{1+}\rangle$ becomes photon-like, i.e.
$|\psi_{1+}\rangle \rightarrow  |1,\downarrow\rangle$, while the state
$|\psi_{1-}\rangle$ becomes qubit-like, i.e., $|\psi_{1-}\rangle\rightarrow
|0,\uparrow\rangle$. For large negative detuning $-\delta\gg g$ with $\theta_n
\rightarrow 0^+$, the situation is reversed: $|\psi_{1+}\rangle\rightarrow
|0,\uparrow\rangle$, $|\psi_{1-}\rangle \rightarrow  |1,\downarrow\rangle$.

In the limit $\delta \gg g$ ($-\delta \gg g$), photons with energies around
$\varepsilon_{1+} (\varepsilon_{1-})$ effectively scatter at a Kerr
non-linearity described by the Hamiltonian $H_{\rm \scriptscriptstyle
Kerr}=\bar{\omega}_c b^\dagger b + (U/2) b^\dagger b^\dagger b b$ with energy
$\bar{\omega}_c\approx\omega_c+g^2/\delta - g^4/\delta^3$ and a weak
non-linearity $U\approx-2g^{4}/\delta^3$, where the polaronic shift in the
energy as well as the non-linearity are induced by the presence of the
two-level system. Note that sign$(U)$ is opposite to sign$(\delta)$. On the
other hand, photons with energies around $\varepsilon_{1-} (\varepsilon_{1+})$
scatter mostly at the weakly coupled two-level system with transition
frequency $\bar{\omega}_q\approx\omega_q-g^2/\delta +g^4/\delta^3$, where the
weak coupling gives rise to a small width $\bar{\kappa}\approx\kappa
g^2/\delta^2$.

The single-photon and two-photon scattering matrices of the JC nonlinearity
given by Eqs.~\eqref{eq:r}, \eqref{eq:t}, and \eqref{tmelement} simplify to
the scattering matrices of a Kerr nonlinearity \cite{liao:10} resp. a
two-level system \cite{shen:07} (TLS) in the corresponding energy ranges
($2\nu_0\approx 2\omega_q$ resp. $2\nu_0\approx 2\omega_c$) at large detuning,
as shown in App.~\ref{app:kerr_TLS}. Making use of these scattering matrices,
we may calculate the HOM coefficient $\gamma$ from Eq.~\eqref{eq:gammaHOMgen}
in the Kerr regime, yielding
\begin{align}
\gamma&=
1-\frac{(2\kappa)^2U [2 e_{0c}\xi +U(2\kappa+3 \xi)]}{\xi  [e_{0c}^2+(2\kappa+3 \xi)^2] \left(U^2+4 \xi ^2\right)}\nonumber\\
&+\frac{2 (2\kappa)^2 e_{0c}\xi -[2 \xi ^2+(2\kappa)\xi -(2\kappa)^2] (2\kappa) U}{\xi  U [ e_{0c}^2+(\xi +(2\kappa))^2]}
\nonumber\\
&\hspace{20pt}
+\frac{4 \xi (2\kappa)^2 [(\xi -(2\kappa)) U-2 e_{0c}\xi ]}{U [U^2+4 \xi ^2] [(e_{0c}-U)^2+(\xi +2\kappa)^2]},
\end{align}
with $e_{0c}=\nu_{01}+\nu_{02}-2\bar{\omega}_c$. In the TLS regime, we obtain
\begin{align}
\gamma=&\,1
+\frac{-2i\bar{\kappa}}{e_{0q}+i(2\bar{\kappa}+\xi)}
+\frac{(-2i\bar{\kappa})^2}{(e_{0q}+i(2\bar{\kappa}+\xi))(e_{0q}+i(2\bar{\kappa}+3\xi))}
\nonumber\\
&+\frac{2i\bar{\kappa}}{e_{0q}-i(2\bar{\kappa}+\xi)}
+\frac{(2i\bar{\kappa})^2}{(e_{0q}-i(2\bar{\kappa}+\xi))(e_{0q}-i(2\bar{\kappa}+3\xi))}
\end{align}
with $e_{0q}=\nu_{01}+\nu_{02}-2\bar{\omega}_q$. These simple expressions show
good agreement with the HOM features in the dispersive limit in the
corresponding energy regimes.

\section{Correlations}

The HOM parameter $\gamma$ discussed in the previous section is rather
difficult to measure directly. Instead, it is more convenient to study the
second-order correlation function
\begin{align}
\label{g2ij}
G^{(2)}_{ij}(\tau)&=
\!\!\int \!\!dt \,\!
\langle
a_{ix}^\dagger(t) a_{jx}^\dagger(t+\tau) a_{jx}(t+\tau) a_{ix}(t)
 \rangle
\end{align}
where %
\begin{align}
a_{ix}(t)=\frac{1}{\sqrt{2\pi}} \int d\nu e^{-i\nu (t-x/v)}a_{i\nu},
\end{align}
represents the photon annihilation operator at a particular position
$x\rightarrow +\infty$ in waveguide $i$ (such that $G^{(2)}_{ij}(\tau)$ is
independent of position) and $v$ is the photon group velocity.  Here and in
the following, expectation values are calculated with respect to the outgoing
state $\langle\cdot\rangle=\langle\psi_{\rm out}|\cdot|\psi_{\rm
out}\rangle$.\\

It is straightforward to show that the cross correlation function
$G^{(2)}_{12}(\tau)$ is related to the Hong-Ou-Mandel parameter by simple
integration, i.e.,
\begin{align}
\label{n1n2relation}
 \int d\tau\, G^{(2)}_{12}(\tau)= \langle n_1 n_2 \rangle =\gamma\,.
\end{align}
For a perfect HOM interference, both photons end up in the same waveguide and
thus the cross correlation function vanishes altogether.

By integrating the auto-correlation function $G^{(2)}_{11}(\tau)$ over time,
one obtains the difference between the second and the first moment of the
photon number distribution, i.e.,
\begin{align}
\label{eq:secmomg11}
\int d\tau G^{(2)}_{11}(\tau) = \langle n_1\left(n_1 -1 \right) \rangle
=
 \langle n_1^2 \rangle- \langle n_1 \rangle.
\end{align}
In the special case of two indistinguishable photons with zero time delay and
zero energy detuning, there is a perfect symmetry between waveguide 1 and 2
which leads to $\langle n_1\rangle=\langle n_2\rangle=1$ and $\langle
n_1^2\rangle=\langle n_2^2\rangle$. Furthermore, making use of $\langle
(n_1+n_2)^2\rangle=4$, we find $\langle n_1^2\rangle=2-\gamma$ and
correspondingly
\begin{align}
\label{eq:var}
\langle \Delta n_1^2\rangle=\langle n_1^2\rangle-\langle n_1\rangle^2=1-\gamma.
\end{align}
Note that the expressions in \eqref{eq:secmomg11} and \eqref{eq:var} are
identical in this special case. Thus, in this case, perfect HOM interference
($\gamma=0$) is associated with a Poissonian photon number distribution
characterized by $\langle \Delta n_1^2\rangle=\langle n_1\rangle$, while any
deviations from the ideal HOM conditions lead to sub-poissonian statistics
with $\langle \Delta n_1^2\rangle<\langle n_1\rangle$ (no super-poissonian
statistics is possible for the two photon scattering considered here). The
time-integrated auto-correlation function thus also serves as a useful measure
for the degree of path entanglement generated by the HOM effect.\\

Making use of the outgoing state Eq.~\eqref{eq:outstate}, we find
\begin{align}
\label{g12calc}
G^{(2)}_{ij}(\tau)=\frac{1}{2\pi}\int dE\,
\Bigl|
\int d\Delta e^{i\Delta \tau}\A^{ij}_{E/2+\Delta,E/2-\Delta}\Bigr|^2
\end{align}
with
$\mathcal{A}^{ij}_{\nu^\prime_1,\nu^\prime_2}=
\mathcal{A}^{ij,\rm lin}_{\nu^\prime_1,\nu^\prime_2} +\mathcal{C}_{\nu^\prime_1,\nu^\prime_2}$, where
\begin{align}
\mathcal{A}^{11,\rm lin}_{\nu^\prime_1,\nu^\prime_2}=
r_{\nu_1^\prime}t_{\nu_2^\prime}f^{\rs (1)}_{\nu_1^\prime}f^{\rs (2)}_{\nu_2^\prime}
+t_{\nu_1^\prime} r_{\nu_2^\prime}f^{\rs (1)}_{\nu_2^\prime}f^{\rs (2)}_{\nu_1^\prime}
\end{align}
and $\mathcal{A}^{12,\rm lin}_{\nu^\prime_1,\nu^\prime_2}$ and
$\mathcal{C}_{\nu^\prime_1,\nu^\prime_2}$ defined in Eqs.~(\ref{Alin}) and
\eqref{tcontrib}. The expression \eqref{g12calc} can again be evaluated
analytically. In Fig.~\ref{fig:corrHOM} and Fig.~\ref{fig:corrHOM5} we show
the normalized second-order correlation functions
\begin{align}
g^{(2)}_{ij}(\tau)=\frac{G^{(2)}_{ij}(\tau)}{G^{(2)}_{ij,\infty}}\,,
\label{normcross}
\end{align}
where
\begin{align}
\label{normcorr}
G^{(2)}_{ij,\infty}=
\lim_{\Delta t\rightarrow \infty} \left[ G^{(2)}_{ij}(\tau=\Delta t) 
+ G^{(2)}_{ij}(\tau=-\Delta t)\right]
\end{align}
denotes the uncorrelated contribution to $G^{(2)}_{ij}$ obtained from an
incoming state with a large time delay $\Delta t \gg 1/\xi$ between the two
wave packets describing the case of independent scattering of two
distinguishable (classical) particles.\\ The correlation functions
$g^{(2)}_{ij}(\tau)$ are affected by both, the statistical nature of the
photons as well as correlation effects induced by the non-linearity. In the
following, we will first describe the signatures of the HOM effect in the
cross correlation function and then discuss correlation effects due to the
Jaynes-Cummings nonlinearity.
\subsection{Signatures of HOM interference}
\begin{figure}[t]
\begin{center}
\includegraphics[width=0.49\textwidth,clip]{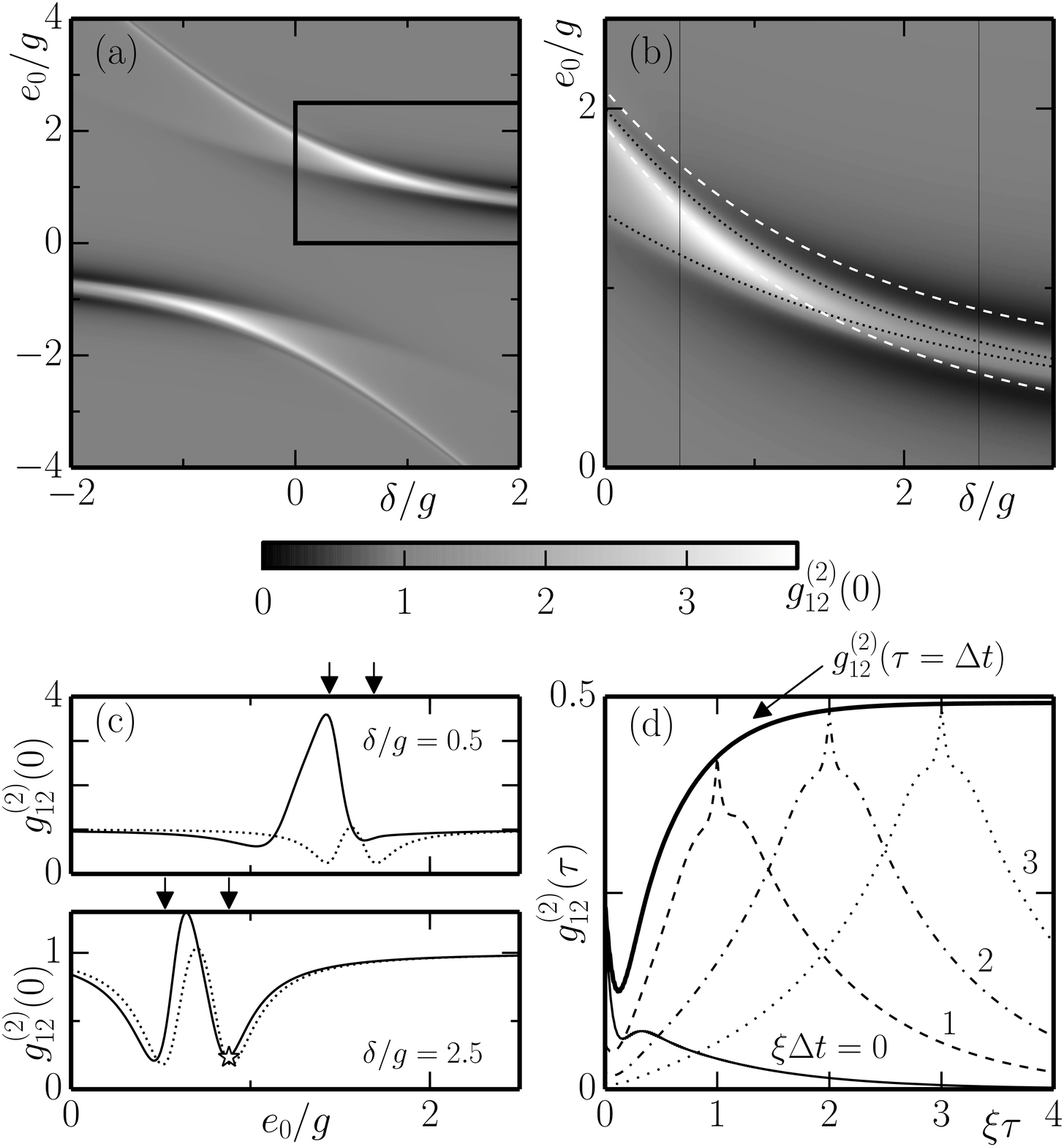}
\caption{
\label{fig:corrHOM}
(a) Normalized cross correlation at zero time delay  $g^{(2)}_{12}(0)$ as
defined in Eq.~(\ref{normcross}) as a function of detuning $\delta$ and
two-photon energy $e_0=2\nu_0-2\omega_c$ as defined in the previous figure
caption.  (b) Inset in (a) (see black box) together with the two photon
energies depicted in Fig.~\ref{fig:fig2new}(a) (dashed and dotted lines with
the same notation as in Fig.~\ref{fig:fig2new}(a)).  (c)~Cuts through (b) at
constant detuning (indicated with vertical lines in (b)) for wave packets of
width $\xi/\kappa=0.1$. The dotted lines correspond to the linear
approximation, where correlation contributions due to the $T$-matrix elements
in Eq.~(\ref{tcontrib}) are neglected. The arrows correspond to
$2\Omega^{(1,2)}_+$. (d) Time-dependence of the normalized second-order cross
correlation function for the parameter values indicated by the star in the
lower panel of (c), i.e., for almost ideal HOM conditions, but for different
time delay $\Delta t$ of the initial single-photon wave packets. The cross
correlation function displays characteristic side peaks at time delay $\Delta
t$, whose height, i.e. $g^{(2)}_{12}( \tau=\Delta t )$  converges at large
time delays ($\Delta t\gg 1/\xi \gg 1/\kappa $), i.e. for completely
distinguishable photon wave packets, towards $1/2$ (see text for explanation).
}
\end{center}
\end{figure}

Figures \ref{fig:corrHOM}(a) and (b) show that the HOM effect at
$2\nu_0\approx 2\Omega^{(1,2)}_\sigma$ (dashed lines) manifests itself as a
suppression of the cross correlations at zero time-delay with $g^{(2)}_{12}(0)
\approx 0$ only in the dispersive regime ($|\delta| \gg g$) for upper
polaritons at positive detuning ($\delta > 0$) and lower polaritons at
negative detuning ($\delta < 0$), where nonlinear effects are weak and the JC
nonlinearity acts as an almost ideal beam splitter.  The different behavior at
small and large detuning is analyzed in more detail in Figures
\ref{fig:corrHOM}(c) showing cuts through the inset in \ref{fig:corrHOM}(b)
(indicated by vertical lines): strong deviations from the linear result
(dotted lines) are observed for small detuning (upper panel), but can be
neglected for $|\delta| \gg g$ (lower panel). The zero-time delay cross
correlation function $g^{(2)}_{12}(0)$ thus provides a useful signature for
path entanglement only in the dispersive regime. Note that for an ideal HOM
effect in the dispersive regime with $\gamma=0$ the wave-packet width should
also tend to zero. The nonlinear effects at small detuning will be discussed
in the next subsection further below.

The quantum interference process leading to the HOM effect relies on the
indistinguishability of the two photons at zero time delay and equal energies.
Thus, a time delay between the incoming photons leads to distinguishability
and suppresses the HOM interference.  Figure \ref{fig:corrHOM} (d) shows the
cross correlation function with a time delay $\Delta t$ of the two photon wave
packets in the dispersive regime for parameter settings corresponding to the
star in the lower panel of Figs. \ref{fig:corrHOM}(c) (HOM condition). We
observe broad peaks centered at the time delay $\tau\approx \Delta t$, where
the first photon has maximal overlap with the second. The overall width
$\sim1/\xi$ of the side peak is given by the width of the wave packet while
the cusp like feature on top arises due to the coupling to the waveguides
$\sim 1/\kappa$ and is suppressed in the limit $\xi/\kappa \rightarrow 0$. For
large time delay and sharp wave packets such that $\Delta t\gg 1/\xi \gg
1/\kappa $, the cross correlation function is given by
\begin{align} 
G^{(2)}_{12}(\tau) \approx \frac{\xi}{8}
\left( e^{-\xi |\tau-\Delta t|} + e^{-\xi |\tau+\Delta t|}\right)\,,
\end{align} 
which yields the asymptotic value $g^{(2)}_{12}( \tau=\Delta t ) \approx 1/2$
in agreement with Figure \ref{fig:corrHOM}(d). The asymptotic value of $1/2$
originates from the fact, that only one of the two classical processes
contributes to $G^{(2)}_{12}(\tau=\Delta t)$ while the normalisation in
Eq.~(\ref{normcorr}) is obtained by summing over both classical paths.

The appearance of a maximum in the cross correlation function at finite times
thus serves as a measure for the distinguishability of the two photons due to
an initial time delay and can be used to benchmark the two single photon
sources attached to both transmission lines. These findings, valid in the
dispersive regime, are consistent with recent experimental circuit QED results
for a static, linear $50/50$ beam splitter in
Refs.~\cite{lang2013,woolley2013}.

\subsection{Signatures of photon nonlinearities}

In the previous subsection, we mostly discussed signatures of the HOM effect
in the dispersive regime, where nonlinear effects can be neglected. We will
now focus on a discussion of the resonant regime, where photon nonlinearities
are strong.\\

\begin{figure}[t]
\begin{center}
\includegraphics[width=0.5\textwidth,clip]{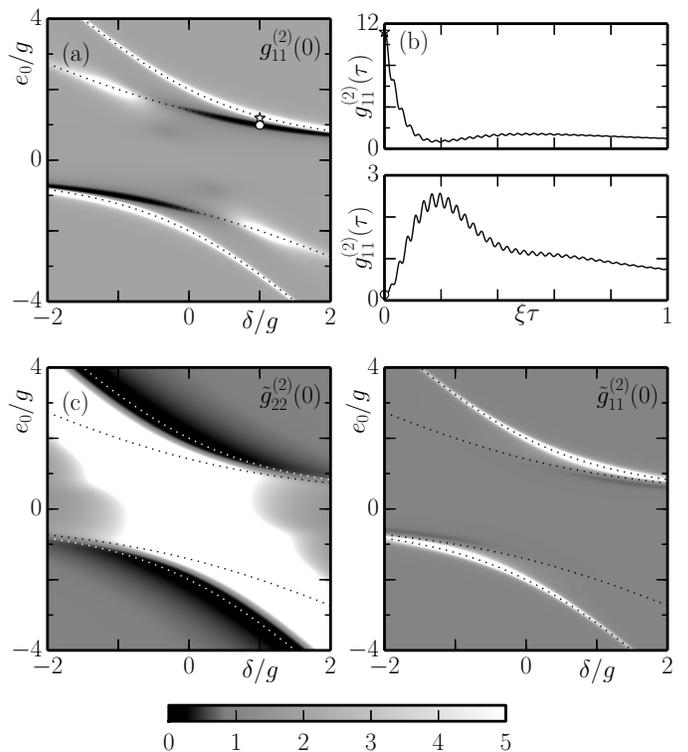}
\caption{
(a) Normalized auto-correlation function $g^{(2)}_{11}(0)$ at zero time delay
as defined in Eq.~(\ref{normcross}) as a function of detuning $\delta$ and
two-photon energy $e_0=2\nu_0-2\omega_c$ for two photons incoming from
different waveguides ($\xi/\kappa=0.1$).  (b) Time-dependence of the
auto-correlation function $g^{(2)}_{11}(\tau)$ for the case of bunching (upper
panel, indicated by star in (a)) and anti-bunching (lower panel, indicated by
circle in (a)).  (c) Normalized correlation function of transmitted (left) and
reflected (right) photons $\tilde{g}^{(2)}_{22}(0)$ and
$\tilde{g}^{(2)}_{11}(0)$ at zero time delay for two photons incoming from the
same waveguide 1. Here, the dotted lines indicate the resonances at
$2\varepsilon_{1,+}$ and $\varepsilon_{2,+}$, with the same notation as in
Fig.~\ref{fig:fig2new}(a). Note, that the correlation functions are cut-off at
an arbitrarily chosen value $\tilde{g}^{(2)}_{ii}(0) < 5$.
}
\label{fig:corrHOM5}
\end{center}
\end{figure}

When both photons are resonant with the one-polariton state
($2\nu_0=2\varepsilon_{1,\pm}$), dipole-induced trans\-parency with reflection
amplitude $r_{\nu=\varepsilon_{1,\pm}}=0$ (cf. Fig.~\ref{fig:fig2new} (b))
would lead to complete and independent transmission of the two particles in
the absence of any correlations, i.e., if one neglects the contributions of
the $T$-matrix in Eq.~(\ref{g12calc}).  However, when qubit and cavity are on
resonance ($\delta\approx 0$), those contributions dominate and lead to photon
blockade. In that case, a single photon already present in the cavity blocks
the transmission of a second photon, which is thus reflected and
preferentially drags the cavity-photon into the same waveguide leading to
bunched correlations with $g_{11}^{(2)}(0)\gg 1$ (see
Fig.~\ref{fig:corrHOM5}(a)).  Interestingly, when the total energy of both
photons matches the two-polariton state energy ($2\nu_0=\varepsilon_{2,\pm}$),
bunched as well as anti-bunched correlations can be observed, depending on the
value of the detuning parameter \footnote{It would be worthwhile to
investigate whether this behaviour also persists in the case of
continuous-wave driving, which is, however, beyond the scope of this
paper.}.\\

Note, that this behaviour should be contrasted with the case, where both
photons impinge on the qubit-cavity system from the same waveguide, i.e., for
an incoming state $|\psi_{\rm in}\rangle\sim\iint d\nu_1  d\nu_2 f^{\rs
(1)}_{\nu_1} f^{\rs (2)}_{\nu_2} a^\dagger_{1\nu_1}a^\dagger_{1\nu_2}| 0
\rangle$. This case is shown in Fig.~\ref{fig:corrHOM5}(c), where photon
blockade at $2\nu_0=2\varepsilon_{1,\pm}$ leads to bunching in reflection
(right panel), but anti-bunching in transmission (left panel). In the latter
case, bunching correlations are found for energies lying between the two
anti-bunched regions. These findings are in agreement with previous
theoretical \cite{shi:11} as well as experimental studies \cite{reinhard2012}.

\section{Conclusion}

In this work, we have studied in detail the interplay of quantum interference
leading to the HOM effect and effective photon-photon interactions in a
wave\-guide QED system, where two single photon wave packets are incident on a
Jaynes-Cummings nonlinearity from two separate waveguides. For this purpose,
we have calculated the cross and auto-correlation functions of the two
wave\-guides analytically based on a scattering matrix approach.  A central
result of our study is that the proposed setup can be used as an in-situ
tunable HOM interferometer in the dispersive regime for already moderate
detuning between the two-level system and the cavity and rather sharp wave
packets of width $\xi$ smaller than the cavity decay width $\kappa$. In the
opposite regime, where two-level system and cavity are on resonance, HOM
interference is washed out and signatures of photon blockade due to effective
photon-photon interactions induced by strong coupling of photons and two-level
system, manifest as bunched correlations in the second-order auto-correlation
function.

\appendix

\section{Derivation of the scattering matrix}\label{app:sca-ma}

\subsection{Even/odd modes}

To calculate the one and two-photon scattering matrix in (\ref{spmatrix}) and
(\ref{tpmatrix}), it is convenient to first decompose the modes of waveguides
$1$ and $2$ into even and odd modes \cite{shen:07,shi:09} by introducing
\begin{eqnarray}
a^{(e,o)}_{\nu}=(a_{1\nu} \pm a_{2\nu})/\sqrt{2}
\end{eqnarray} 
and rewriting the waveguide Hamiltonian as $H_{\rm w}=H_{\rm w}^{(e)}+H_{\rm
w}^{(o)}$ with $H_{\rm w}^{(e,o)}=\int d\nu\, \nu\,
a^{(e,o)\dagger}_{\nu}a^{(e,o)}_{\nu}$.  With this transformation only the
even modes couple to the cavity via
\begin{eqnarray}
\label{evenint}
V=\sqrt{2} g_\mathrm{w}\int d\nu\,(b^\dagger a^{(e)}_{\nu}+b a^{(e)\dagger}_{\nu})\,, 
\end{eqnarray}
and thus the Hamiltonian for the odd modes becomes trivial. The scattering
matrix associated with the odd modes is given simply by the identity matrix.
The Hamiltonian in the even modes,
\begin{eqnarray}
\label{evenham}
H^{(e)}=H^{(e)}_{\rm w}+H_{\rm\scriptscriptstyle JC} +V,
\end{eqnarray}
then describes one chiral mode linearly coupled to the cavity.  The scattering
matrix for the even modes is nontrivial and can be calculated using the
formalism in Ref.~\onlinecite{pletyukhov2012} (see next section).

Once we have calculated the one photon scattering matrices in the even
subspace, i.e., $S^{(e)}_{\nu^\prime\nu}=\langle 0 |
a^{(e)}_{\nu^\prime}Sa^{(e)\dagger}_{\nu} |0\rangle$, we transform back to
physical space via the relations
\begin{eqnarray}
S^{11}_{\nu^\prime\nu}&=&S^{22}_{\nu^\prime\nu}=\frac{1}{2}(S^{(e)}_{\nu^\prime\nu}+\delta_{\nu^\prime\nu})\\
S^{12}_{\nu^\prime\nu}&=&S^{21}_{\nu^\prime\nu}=\frac{1}{2}(S^{(e)}_{\nu^\prime\nu}-\delta_{\nu^\prime\nu})\,,
\end{eqnarray}
where the delta function on the right hand side stems from the trivial
contribution of the odd modes.  From the two-photon scattering matrix in the
even space $S^{(e)}_{\nu^\prime_1\nu^\prime_2\nu_1 \nu_2}=\langle 0|
a^{(e)}_{\nu^\prime_{2}}a^{(e)}_{\nu^\prime_{1}}Sa^{(e)\dagger}_{\nu_1}a^{(e)\dagger}_{\nu_2}
|0\rangle=S^{(e)}_{\nu^\prime_1\nu_1}S^{(e)}_{\nu^\prime_2\nu_2}+
S^{(e)}_{\nu^\prime_1\nu_2}S^{(e)}_{\nu^\prime_2\nu_1}+i\T^{(2e)}_{\nu^\prime_1\nu^\prime_2,\nu_1\nu_2}$
we obtain the two-photon scattering matrix in Eq.~(\ref{tpmatrix}) from the
results above together with the simple relation
\begin{eqnarray}
\T^{(2)}_{\nu^\prime_1\nu^\prime_2,\nu_1\nu_2}=\T^{(2e)}_{\nu^\prime_1\nu^\prime_2,\nu_1\nu_2}/4\,.
\end{eqnarray}

\subsection{$T$-matrix}
To calculate the single- and two-photon scattering matrix of the even modes we
make use of the formalism introduced in Ref.~\onlinecite{pletyukhov2012}.  The
scattering matrix is conveniently expressed through the $T$-matrix which
contains the non-trivial part of the scattering, i.e.,
\begin{align}
S=1-2\pi i \,\delta(E_{\rm in}-E_{\rm out}) T(E_{\rm in})
\end{align}
with the energies $E_{\rm in}$ and $E_{\rm out}$ of the incoming resp.
outgoing state.  The $T$-matrix can be expressed through the full Green's
function $\hat{G}(\omega)=(\omega-H+i0^+)^{-1}$ with $H^{(e)}$ defined in
Eq.~\eqref{evenham} as
\begin{align}
T(\omega)=V+V \hat{G}(\omega) V\,.
\end{align}
The series representation of the full Green's function $\hat{G}(\omega)=
\sum_{n=0}^\infty G_0(\omega)\left[ VG_0(\omega)\right]^n$ in terms of the
free Green's function $G_0(\omega)=(\omega-H_0+i0^+)^{-1}$ and the interaction
$V$ given in Eq.~\eqref{evenint} yields the corresponding series
representation of the $T$-matrix.  For photon number conserving scattering
processes we obtain $T(\omega)=V \sum_{n=1}^\infty
\left[G_0(\omega)V\right]^n$.  The main result of
Ref.~\onlinecite{pletyukhov2012} is to show that the $T$-matrix for an
incoming $N$-photon state can be expressed as
\begin{align}
\label{resultT}
T^{(N)}(\omega) = \vdots \, G_0^{-1} \tilde{G} \left(V\tilde{G}\right)^{2N} G_0^{-1} \, \vdots \, , 
\end{align}
with the dressed Green's function 
\begin{align}
\label{dressedG}
\tilde{G}(\omega)=(\omega-H_0-\Sigma)^{-1}\,,
\end{align}
where the operator $\Sigma$ (of the form of a self-energy) accounts for the
coupling of the atomic system to the waveguide and for the linear coupling in
Eq.~\eqref{evenint} attains the remarkably simple form
\begin{eqnarray}
\Sigma= -2i\pi g_\mathrm{w}^2 b^\dagger b\,.
\end{eqnarray}
The operation $\vdots ( \,\, ) \vdots $ is a version of the normal ordering,
which removes contractions in pairs of $V$'s, but does not account for
contractions between $V$ and $H_0$. The latter can be effectively accounted
for by shifts in $\omega$-arguments of the dressed Green's functions
$\tilde{G}$ occurring in the expansion \eqref{resultT}.

\subsection{Scattering matrix for even modes}
We are now in the position to calculate, e.g., the one photon scattering matrix elements
\begin{eqnarray}
S^{(e)}_{\nu^\prime\nu}&=&\langle 0 | a^{(e)}_{\nu^\prime}Sa^{(e)\dagger}_{\nu} |0\rangle
= \delta(\nu'-\nu)\\ 
&&-2\pi i \delta(\nu'-\nu) \langle 0| a^{(e)}_{\nu'}T^{(1)}(\nu)a^{(e)\dagger}_{\nu}|0\rangle\nonumber
\end{eqnarray}
from the result in (\ref{resultT}), i.e.,
\begin{align}
\label{resultT1}
T^{(1)}(\omega)= \vdots \, G_{0}^{-1}(\omega)\tilde{G}(\omega)V\tilde{G}(\omega)V\tilde{G}(\omega)G_{0}^{-1}(\omega)\, \vdots \,.
\end{align}
Here, the combination $\tilde{G}(E_{\rm in})G_{0}^{-1}(E_{\rm in})$ at the end
of the operator chain on the r.h.s of Eq.~(\ref{resultT1}) acts as a projector
on the dark states (non-broadened states). This can be seen from the
expression $\tilde{G} (E_{\rm in})G_{0}^{-1}(E_{\rm in})|\psi_{\rm
in}\rangle$, which is zero for any incoming eigenstate of $H_0$ since
$G_0^{-1}(E_{\rm in})=(E_{\rm in}-H_0)$ except when $\tilde{G} (E_{\rm
in})=G_{0}(E_{\rm in})$, i.e., for the non-broadened states with $\Sigma
|\psi_{\rm in}\rangle=0$. In the case of the JC nonlinearity, the only
non-broadened state is the ground state of the combined cavity-qubit system
$|\psi_{0}\rangle=|0,\downarrow\rangle$, such that we can replace
$\tilde{G}(E_{\rm in})G_0^{-1}(E_{\rm in})$ by the projector
$P_0=|\psi_{0}\rangle\langle\psi_{0}|$. The same is true for
$G_{0}^{-1}(E_{\rm in})\tilde{G} (E_{\rm in})$ on the left of the expression
(as $E_{\rm in}=E_{\rm out}$).  By expressing the dressed Greens function in
(\ref{dressedG}) through the projector
$\tilde{P}_{n\sigma}=|\tilde{\psi}_{n\sigma}\rangle \langle
\tilde{\psi}_{n\sigma}|$ on the Jaynes-Cummings resonances
$|\tilde{\psi}_{n\sigma}\rangle$ (given by the expressions for
$|\psi_{n\sigma}\rangle$ in (\ref{jcstates}) with the angle $\theta_{n}$
replaced by $\tilde{\theta}_{n}$ with
$\omega_c\rightarrow\tilde{\omega}_c=\omega_c-i \kappa$) as $\tilde{G}
(\omega)=\sum_{n\alpha}\tilde{P}_{n\alpha}(\omega-H_{\rm
w}-\tilde{\varepsilon}_{n\alpha})^{-1}$ and calculating the corresponding
matrix elements in (\ref{resultT1}), one arrives at the final result
$S_{\nu'\nu}^{(e)} = s_{\nu}^{(e)} \delta (\nu'-\nu)$ with
\begin{align}
s_{\nu}^{(e)}=\frac{(\nu-\tilde{\varepsilon}_{1+}^\ast)(\nu-\tilde{\varepsilon}_{1-}^\ast)}
{(\nu-\tilde{\varepsilon}_{1+})(\nu-\tilde{\varepsilon}_{1-})}.
\label{eq:se}
\end{align}
A similar calculation for the two-photon $T$-matrix
\begin{eqnarray}
T^{(2)}(\omega)= \vdots \, P_{0}V \tilde{G} (\omega)V \tilde{G} (\omega)V \tilde{G} (\omega)VP_{0} \, \vdots
\end{eqnarray}
yields after some lengthy algebra the result stated in Eq.~(\ref{tmelement}).

\section{Strongly dispersive regime: Limit of Kerr nonlinearity and TLS scattering matrix}\label{app:kerr_TLS}

We show here, that in the dispersive limit $\delta\gg g$ the JC scattering
matrix given by Eqs.~\eqref{eq:r}, \eqref{eq:t}, and \eqref{tmelement} can be
simplified to the scattering matrix of the Kerr nonlinearity for incoming
photon energies $\nu_{0i}\approx \omega_c$ and to the scattering matrix of a
TLS for energies $\nu_{0i}\approx \omega_q$. [The limit $-\delta \gg g$ can be
treated analogously.]

Let us start with the Kerr regime: We approximate the
energies~\eqref{jcenergies} as $\varepsilon_{n+}=\omega_c+ng^2/\delta-n^2
g^4/\delta^3$, such that the two lowest levels effectively form a Kerr
nonlinearity with $\varepsilon_{1+}=\bar{\omega}_c$ and
$\varepsilon_{2+}=2\bar{\omega}_c+U$ with
$\bar{\omega}_c\approx\omega_c+g^2/\delta - g^4/\delta^3$ and $U\approx
-2g^4/\delta^3$. The energies of the off-resonant states $|\psi_{n-}\rangle$
are approximated as $\varepsilon_{n-}\approx n\omega_c-\delta$. To approximate
the single-photon scattering matrix coefficients $t_{\nu}$ and $r_{\nu}$ given
by Eqs.~\eqref{eq:r} and \eqref{eq:t}, it is useful to note that they can be
written as $r_{\nu}=(s_{\nu}^{(e)}+1)/2$ and $t_{\nu}=(s_{\nu}^{(e)}-1)/2$.
As $\tilde{\varepsilon}_{1-}$ is off-resonant, we approximate
$\nu-\tilde{\varepsilon}_{1-} \approx \varepsilon_{1+}
-\tilde{\varepsilon}_{1-} \approx \delta$ which leads to
$s_{\nu}^{(e)}=(\nu-\bar{\omega}_c-i\kappa)/(\nu-\bar{\omega}_c+i\kappa)$ in
agreement with Ref.~\onlinecite{liao:10} and to
$r_{\nu}=(\nu-\bar{\omega}_c)/(\nu-\bar{\omega}_c+i\kappa)$ and
$t_{\nu}=-i\kappa/(\nu-\bar{\omega}_c+i\kappa)$.  In the same way, we
approximate the $T$-matrix given by Eq.~\eqref{tmelement}: Only the
single-photon resonances at energy $\varepsilon_{1+}$, the two photon
resonances at energies $\varepsilon_{2+}$, as well as the energy dependence in
$\nu_1+\nu_2-2\varepsilon_{1+}$ are relevant. In all the other factors which
do not contain $\varepsilon_{n+}$ we approximate $\nu_{i} \approx
\varepsilon_{1+}$ and obtain to leading order in $g/\delta$
\begin{align}
\mathcal{T}^{(2)}_{\nu^\prime_1\nu^\prime_2,\nu_1 \nu_2}
&\approx-
\frac{\kappa^2U}{\pi}
\frac{\nu_1+\nu_2-2\bar{\omega}_c+2i\kappa}{\nu_1+\nu_2-2\bar{\omega}_c-U+2i\kappa}\\
&\quad\times\frac{\delta_{\nu^\prime_1+\nu^\prime_2,\nu_1+\nu_2}}{\prod_{i=1,2}(\nu^\prime_i-\bar{\omega}_c+i\kappa)
(\nu_i-\bar{\omega}_c+i\kappa)},\nonumber
\end{align}
with $\bar{\omega}_c$ and $U$ as introduced above.

For incoming photon energies $\nu_{0i}$ close to $\varepsilon_{1-}$, we
proceed similarly. In this regime, only the resonance at
$\tilde{\varepsilon}_{1-}$ is relevant which we approximate to leading order
in $g/\delta$ by $\tilde{\varepsilon}_{1-}=\bar{\omega}_q-i\bar{\kappa}$ with
$\bar{\omega}_q=\omega_q-g^2/\delta$ and $\bar{\kappa}=\kappa g^2/\delta^2$.
All the other factors which do not contain $\tilde{\varepsilon}_{1-}$ are
off-resonant and approximated by
$\tilde{\varepsilon}_{n+}\approx\varepsilon_{n+}\approx n\omega_c$ and
$\tilde{\varepsilon}_{n-}\approx\varepsilon_{n-}\approx n\omega_c-\delta$ for
$n>1$. Approximating $\nu-\tilde{\varepsilon}_{1+}\approx-\delta$ in
Eq.~\eqref{eq:se}, we obtain
$s_{\nu}^{(e)}=(\nu-\bar{\omega}_q-i\bar{\kappa})/(\nu-\bar{\omega}_q+i\bar{\kappa})$,
in agreement with Ref.~\onlinecite{shen:07}. Similarly as above, we
approximate $\nu_i$ in all terms by $\omega_q$, except for the single photon
resonance at $\tilde{\varepsilon}_{1-}$ and in the factor
$\nu_1+\nu_2-2\tilde{\varepsilon}_{1-}$, and obtain to leading order in
$g/\delta$
\begin{align}
\mathcal{T}^{(2)}_{\nu^\prime_1\nu^\prime_2,\nu_1 \nu_2}
&\approx
\frac{\bar{\kappa}^2}{\pi}
\frac{\delta_{\nu^\prime_1+\nu^\prime_2,\nu_1+\nu_2}(\nu_1+\nu_2-2\bar{\omega}_q+2i\bar{\kappa})}{\prod_{i=1,2}(\nu^\prime_i-\bar{\omega}_q+i\bar{\kappa})(\nu_i-\bar{\omega}_q+i\bar{\kappa})},
\end{align}
in agreement with the result for the TLS in Ref.~\onlinecite{shen:07}.

\end{document}